\documentclass[12pt]{iopart}

\usepackage{harvard}

\usepackage{algorithmic}
\usepackage{graphicx}
\usepackage{textcomp}

\usepackage{enumerate}

\usepackage{algorithm}
\usepackage{algorithmic}

\usepackage{float}
\usepackage{booktabs}
\usepackage{multirow}

\begin{document}

\title[Scatter correction based on quasi-Monte Carlo for CT reconstruction]{Scatter correction based on quasi-Monte Carlo for CT reconstruction}

\author{Guiyuan Lin$^{1}$, Shiwo Deng$^{2,}$$^{*}$, Xiaoqun Wang$^{3}$ and Xing Zhao$^{4}$}

\address{$^{1}$ School of Mathematics and Statistics, Hunan First Normal University, Changsha, 410205, China\\
	$^{2}$ School of Mathematics and Statistics, Hunan University of Technology and Business, Changsha 410205, China\\
	$^{3}$ The Department of Mathematical Sciences, Tsinghua University, Beijing, 100084, China\\
	$^{4}$ School of Mathematical Sciences, Capital Normal University, Beijing, 100048, China\\
	$^{*}$ Author to whom any correspondence should be addressed.}
\ead{dengsw0629@163.com}

%

\begin{abstract}
Scatter signals can degrade the contrast and resolution of computed tomography (CT) images and induce artifacts. How to effectively correct scatter signals in CT has always been a focal point of research for researchers. This work presents a new framework for eliminating scatter artifacts in CT. In the framework, the interaction between photons and matter is characterized as a Markov process, and the calculation of the scatter signal intensity in CT is transformed into the computation of a $4n$-dimensional integral, where $n$ is the highest scatter order. Given the low-frequency characteristics of scatter signals in CT, this paper uses the quasi-Monte Carlo (QMC) method combined with forced fixed detection and down sampling to compute the integral. In the reconstruction process, the impact of scatter signals on the X-ray energy spectrum is considered. A scatter-corrected spectrum estimation method is proposed and applied to estimate the X-ray energy spectrum. Based on the Feldkamp-Davis-Kress (FDK) algorithm, a multi-module coupled reconstruction method, referred to as FDK-QMC-BM4D, has been developed to simultaneously eliminate scatter artifacts, beam hardening artifacts, and noise in CT imaging. Finally, the effectiveness of the FDK-QMC-BM4D method is validated in the Shepp-Logan phantom and head. Compared to the widely recognized Monte Carlo method, which is the most accurate method by now for estimating and correcting scatter signals in CT, the FDK-QMC-BM4D method improves the running speed by approximately $102$ times while ensuring accuracy. By integrating the mechanism of FDK-QMC-BM4D, this study offers a novel approach to addressing artifacts in clinical CT.
\end{abstract}

%
\vspace{2pc}
\noindent{\it Keywords}: scatter correction, quasi-Monte Carlo, beam hardening correction, denoising, CT reconstruction


%
%

\section{Introduction}

{V}{arious} artifacts in computed tomography (CT) images, such as geometric artifacts, beam hardening artifacts, ring artifacts, and scatter artifacts, have always been one of the key factors affecting the quality of CT images. How to effectively correct artifacts in CT images has always been a focal point of research in both the academic and industrial fields of CT. Among them, scatter artifacts are one of the most challenging artifacts to correct in CT images.

At present, there have been many studies on estimating and correcting scatter signals in projection data to improve CT image quality. The methods for scatter artifact correction can be divided into three categories: pure hardware methods, combined hardware and software methods, and pure software methods. The principle of pure hardware methods is to reduce the components of scatter photons in the collected CT scan data through various hardware occlusion techniques \cite{2006zhu,2015Ritschl,2017Bier}. For example, in medical multi-slice helical CT, hardware collimation is used to reduce scatter photons. The principle of the combined hardware and software methods is to first estimate the distribution of scatter photons in the collected CT scan data through hardware, and then correct the scatter photons, such as the scatter grid \cite{grid} and the method based on compressed sensing \cite{2018CS}. The principle of the pure software method is to establish a CT imaging model that includes scatter mechanisms and image reconstruction algorithms. For example, scatter kernel superposition (SKS) methods \cite{SKS1988,SKS2010}, Monte Carlo (MC) methods \cite{2004Colijn,2006Zbijewski,2006Kyriakou,2009Poludniowski,Badal2009Accelerating,JiaFast,2012A,gMMC2019,gMMC2020,2024dark}, Acuros CTS\cite{2018Acuros,2018Acurosb}, 
and deep learning methods \cite{2019Scatter1,2020Roser,2021ScatterDL,2021ScatterDLHe} all first simulate scatter signals and then correct the scatter signals for reconstruction. 

Each category of method has its own advantages and disadvantages. In methods involving hardware, the resolution of the reconstructed image may be affected or the scanning time may increase. For instance, the longitudinal resolution of multi-slice CT is lower than the transverse resolution. In pure software methods, the SKS methods deconvolve scatter from projection data by the point spread function generated from the pencil beam. These are approximate scatter correction methods. Although they have advantages such as high computational efficiency and practicality, they also may suffer from inaccurate estimation of the scatter signal. The MC methods are often considered the most accurate approaches for solving radiation transport problems because they can accurately describe the underlying physical interactions between radiation and matter. In the past 15 years, using MC methods to estimate and correct scatter signals in CT has been a hot topic. However, the enormous computational demands of MC methods have prevented their application in clinical practice so far. Acuros CTS calculates the scatter signal by deterministically solving the Linear Boltzmann Transport Equation, and then further corrects it in CT. This method has been researched by researchers at  Varian for nearly a decade and is currently applied clinically for cancer radiotherapy planning. Deep learning methods are also approximate simulation methods, which require prior training of the model using scatter data from phantoms. These methods take less time for scatter correction using the trained model.

Based on the  deficiencies and characteristics of existing methods for estimating and correcting scatter signals in CT, this paper presents a new framework for eliminating scatter artifacts in CT. In the framework, by characterizing the interaction between photons and matter as a Markov process, the calculation of scatter signal intensity in CT is transformed into the computation of a $4n$-dimensional integral, where $n$ is the highest scatter order. Given the low-frequency characteristics of the scatter signal, this paper uses the quasi-Monte Carlo (QMC) method combined with forced fixed detection and down sampling to simulate the scatter signal, i.e., to estimate a $4n$-dimensional integral. QMC method is a deterministic version of MC method, whose fundamental idea is to use low-discrepancy sequences instead of random numbers for simulation \cite{niederreiter:1992,dick:pillichshammer:2010}. We have attempted to use GPU-based QMC to simulate scatter signals. Compared to the MC methods, it has a faster convergence rate, allowing us to achieve high accuracy with fewer photons, and significantly reducing the runtime \cite{2021Lin}. The forced fixed detection enforces the calculation of the probability of photons reaching each fixed detector pixel at every interaction point, which effectively increases the utilization rate of photons \cite{2006Kyriakou,2009Poludniowski,2008Acceleration}. In the reconstruction process of the framework, the impact of scatter signals on the X-ray spectrum is considered. Based on the existing expectation-maximization (EM) algorithm for estimating the X-ray spectrum, an EM spectrum estimation algorithm with scatter correction (EM-SC) is proposed. Existing studies have shown that the EM algorithm is an accurate and robust method for estimating the X-ray energy spectrum when the projection data is scatter-free \cite{2005EM}. Researchers further studied the estimation of the X-ray energy spectrum using the EM algorithm. In 2015, Lee et al. proposed a method to first estimate the scatter signal based on a single scatter model, and then solve for the X-ray energy spectrum using the EM algorithm \cite{2015Lee}. This method has higher accuracy compared to the approach that directly uses the EM algorithm 
without considering scatters. However, this method estimates the scatter signal based on a single scatter model, while existing research indicates that photons are quite likely to interact multiple times with the measured object. Additionally, this method only accounts for Compton scattering and does not consider the impact of Rayleigh scattering. The EM-SC method proposed in this paper addresses the deficiencies in reference \cite{2015Lee}, which stated as ``In this study, coherent scattering was not taken into account $\cdots$ the further verification of accuracy of IMPACT in estimating coherent scattering coefficient is needed. $\cdots$ Although this study is specifically targeted on Compton scattering correction, we will extend the IMPACT algorithm and single scattering model to incorporate coherent scattering and validate the method in the near future.'' Finally, based on the Feldkamp-Davis-Kress (FDK) reconstruction algorithm \cite{1984FDK}, a multi-module coupled reconstruction method, referred to as FDK-QMC-BM4D, has been developed to simultaneously eliminate scatter artifacts, beam hardening artifacts, and noise in CT imaging. The denoising is performed using the method from reference \cite{BM4D2013}.

The main contributions of this study are (i) characterizing the interaction between photons and matter as a Markov process and transforming the calculation of scatter signals in CT into a $4n$-dimensional integral, (ii) using QMC combined with forced fixed detection and down sampling to compute the $4n$-dimensional integral, and (iii) the consideration of the influence of scatter signals on X-ray energy spectrum, proposing a scatter-corrected spectrum estimation method and applying it to estimate the X-ray energy spectrum. Based on the FDK, a multi-module coupled reconstruction method has been developed  to simultaneously eliminate scatter artifacts, beam hardening artifacts, and noise in CT imaging. 

\section{Theory and Methods}
 Figure 1 
depicts the overall workflow of the FDK-QMC-BM4D method, which uses a head as a carrier. Based on the FDK, the FDK-QMC-BM4D method mainly includes scatter signal simulation, spectrum estimation, scatter signal and beam hardening correction, and denoising. The following sections will provide a detailed introduction. 

\begin{figure}[!t]
	\centerline{\includegraphics[width=\columnwidth]{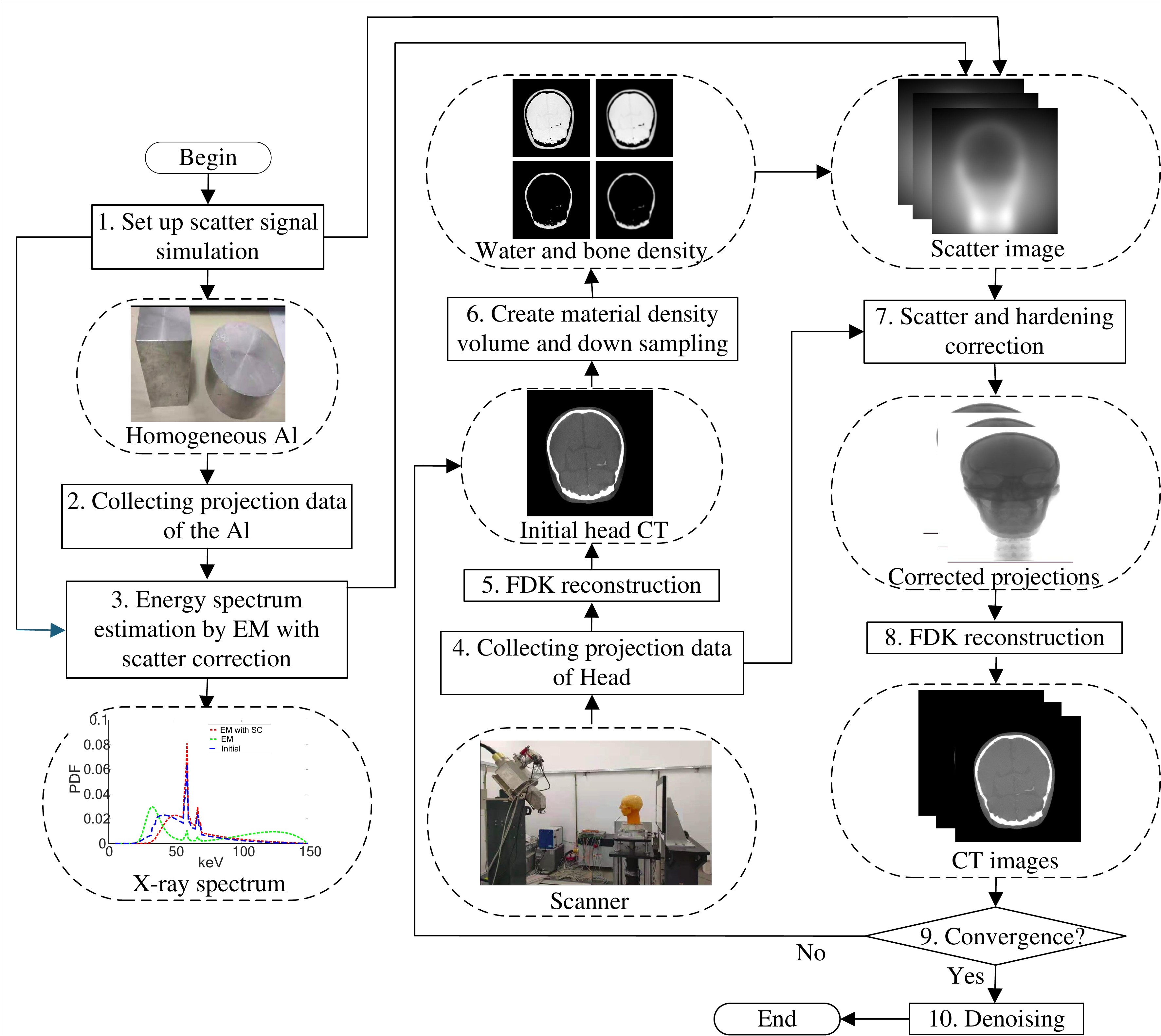}}
	\caption{The overall workflow of the proposed reconstruction method FDK-QMC-BM4D, which uses the head as a carrier.}
	\label{fig:Workflow}
\end{figure}

\subsection{The Scatter Model}
After being emitted from the source S, photons have a certain probability of reaching the measured object M, where they undergo multiple interactions. Theoretically, there is a certain probability that photons will reach each detector pixel from each interaction point. Figure 2 illustrates the geometric paths from each interaction point $A_i(i=1,\cdots,n)$ to the fixed detector pixel $D_j$ after photons interact with the measured object M. Let the path of photons emitted from the X-ray source S, interacting $i$ times with M, and then reaching the fixed detector pixel $D_j$ be denoted as $l_{i,j}:{\rm S}$ $ \rightarrow A_0 \rightarrow A_1 \rightarrow \dots \rightarrow A_i \rightarrow B_{i,j} \rightarrow D_{j}$, $i\in\{1,\dots, n\}, j\in\{1,\dots,m^2\},$ where $m^2$ is the number of detector pixels of D. Let $P(l_{i,j})$ denote the probability of photons traveling along path $l_{i,j}$ reaching detector pixel $D_j$. The transition of the photon from $A_{i-1}$ to $A_i$ is only related to its current interaction point $A_{i-1}$ and is not influenced by the states prior to $A_{i-1}$. Therefore, the interaction between the photon and matter is a Markov process. By combining with forced fixed detection, it can be known that
\begin{eqnarray}\label{plij}
     P(l_{i,j})&=P({\rm S} \rightarrow A_{1}\rightarrow A_{2}\rightarrow\dots\rightarrow A_{i})P(A_{i}
			\rightarrow D_{j})\\&=P({\rm S} \rightarrow A_{1})P(A_{1}\rightarrow A_{2}|{\rm S}\rightarrow A_{1})\\&\prod_{k=3}^{i}P(A_{k-1}\rightarrow A_{k}|A_{k-2}\rightarrow A_{k-1})P(A_{i}
			\rightarrow D_{j}).
\end{eqnarray}
	
\begin{figure}[!t]
	\centerline{\includegraphics[width=\columnwidth]{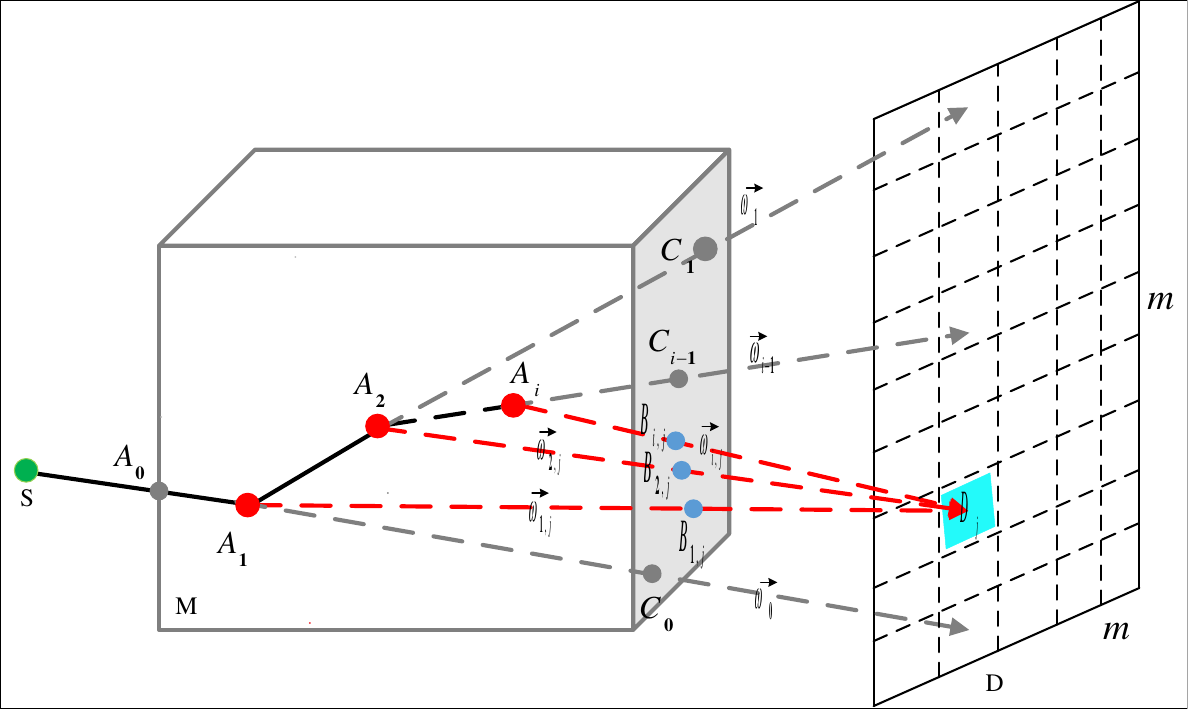}}
	\caption{A simple geometric illustration of photon-matter interaction in CT.}
	\label{fig:fixpathsnew202407}
\end{figure}
	
According to Beer's Law, the probability density function (PDF) for photons traveling from $A_0$ to $A_1$ is
\begin{eqnarray}
	\mu_{tot}(A_1, E_0){\rm exp}\big[{-\int_{0}^{t_1}\mu_{tot}(A_0+t{\mathop{\omega}\limits ^{\rightarrow}}_0, E_0)\mathrm{d}{t}}\big],
\end{eqnarray}
where $\mu_{tot}(\bi{x},E)$ is the linear attenuation coefficient, which is a function of spatial coordinates $\bi{x}$ and the energy $E$. It is the sum of the Compton scattering coefficient $\mu_{T_0}(\bi{x},E)$, the Rayleigh scattering coefficient $\mu_{T_1}(\bi{x},E)$, and the photoelectric effect coefficient $\mu_{a}(\bi{x},E)$. The $A_0$ is the initial intersection point of the photon traveling in the initial unit incident direction ${\mathop{\omega}\limits ^{\rightarrow}}_0$ with M, the $t_1$ is the length of the photon traveling from $A_0$ to $A_1$ along ${\mathop{\omega}\limits ^{\rightarrow}}_0$, and it is a random variable, $A_1=A_0+t_1{\mathop{\omega}\limits ^{\rightarrow}}_0, 0\leq t_1\leq c_0$. The $c_0$ is the length from $A_0$ to $C_0$ along ${\mathop{\omega}\limits ^{\rightarrow}}_0$, the $C_0$ is the intersection point of the photon with M when it escapes M along ${\mathop{\omega}\limits ^{\rightarrow}}_0$ and $C_0=A_0+c_0{\mathop{\omega}\limits ^{\rightarrow}}_0$. The $E_0$ is the initial energy of the photon. Let the initial flux intensity emitted along ${\mathop{\omega}\limits ^{\rightarrow}}_0$ be $I_0({\mathop{\omega}\limits ^{\rightarrow}}_0)$, so
\begin{eqnarray}\label{Pl1}
	   P({\rm S}\rightarrow A_1)&=\int_{4\pi}\mathrm{d}{\mathop{\omega}\limits ^{\rightarrow}}_0\int_{\Omega_{E_0}}\mathrm{d}E_0\int_{0}^{c_0}\mathrm{d}t_1\Big[I_0({\mathop{\omega}\limits ^{\rightarrow}}_0)\phi(E_0)\\&\times \mu_{tot}(A_1,E_0){\rm exp}\big[{-\int_{0}^{t_1}\mu_{tot}(A_0+t{\mathop{\omega}\limits ^{\rightarrow}}_0,E_0)\mathrm{d}t}\big]\Big],
\end{eqnarray} 
where $\phi(E_0)$ is the X-ray energy spectrum, $\Omega_{E_0}$ is the value space of the energy spectrum.
	
The probability of a photon moving from $A_{i-1}$ to $A_i$ is only affected by the energy of the photon at $A_{i-1}$, the type of scatter, and the scatter direction (i.e., it is only influenced by the state of the photon at $A_{i-1}$ and not by any state prior to $A_{i-1}$), thus
\begin{eqnarray}\label{2ii}
	\fl	
		P(A_{i-1}\rightarrow A_{i}|A_{i-2}\rightarrow A_{i-1})&=\sum_{\delta_{i-1}=0}^{1}p_{T_{\delta_{i-1}}}(A_{i-1})\int_{4\pi}\mathrm{d}{\mathop{\omega}\limits ^{\rightarrow}}_{i-1}\int_{0}^{c_{i-1}}\mathrm{d}t_{i}\\
		&\times\Big[p_{\theta_{\delta_{i-1}}}(A_{i-1},E_{i-2}^{\delta_{i-2}}\rightarrow E_{i-1}^{\delta_{i-1}},{\mathop{\omega}\limits ^{\rightarrow}}_{i-2}\rightarrow{\mathop{\omega}\limits ^{\rightarrow}}_{i-1})&\\
		&\times \mu_{tot}(A_{i},E_{i-1}^{\delta_{i-1}}){\rm exp}\big[{-\int_{0}^{t_{i}}\mu_{tot}(A_{i-1}+t{\mathop{\omega}\limits ^{\rightarrow}}_{i-1},E_{i-1}^{\delta_{i-1}})\mathrm{d}t}\big]\Big],&
\end{eqnarray}
where $p_{T_{0}}(A_{i-1})$ and $p_{T_{1}}(A_{i-1})$ represent the probabilities of a photon undergoing Compton scattering and Rayleigh scattering at $A_{i-1}$, respectively. The ${\mathop{\omega}\limits ^{\rightarrow}}_{i-1}$ is the $(i-1)$-th order unit scatter direction of the photon at $A_{i-1}$. If the photon undergoes the next interaction within M, the next-order interaction point will only be on the ray $\overrightarrow{A_{i-1}C_{i-1}}$. Consistent with the $i=1$ case, $C_{i-1}$ is the intersection point with M when the photon escapes M along ${\mathop{\omega}\limits ^{\rightarrow}}_{i-1}$. The $c_{i-1}$ is the length from $A_{i-1}$ to $C_{i-1}$ along ${\mathop{\omega}\limits ^{\rightarrow}}_{i-1}$, $t_{i}$ is the length of the photon traveling from $A_{i-1}$ to $A_{i}$ along ${\mathop{\omega}\limits ^{\rightarrow}}_{i-1}$, satisfying $0\leq t_{i}\leq c_{i-1}$, and it is a random variable. $p_{\theta_{0}}$ and $p_{\theta_{1}}$ are the PDFs of the polar angles for Compton scattering and Rayleigh scattering, respectively, and are the normalized differential cross sections. The
\begin{eqnarray}\label{pscatter1}
\fl		p_{\theta_{0}}(A_{i-1},E_{i-2}^{\delta_{i-2}}\rightarrow E_{i-1}^{0},{\mathop{\omega}\limits ^{\rightarrow}}_{i-2}\rightarrow{\mathop{\omega}\limits ^{\rightarrow}}_{i-1})=\frac{\mu_{T_0}(A_{i-1},E_{i-2}^{\delta_{i-2}}\rightarrow E_{i-1}^{0},{\mathop{\omega}\limits ^{\rightarrow}}_{i-2}\rightarrow{\mathop{\omega}\limits ^{\rightarrow}}_{i-1})}{\mu_{tot}(A_{i-1},E_{i-1}^{0})}	 
\end{eqnarray}
and
\begin{eqnarray}\label{pscatter2}
\fl	 p_{\theta_{1}}(A_{i-1},E_{i-2}^{\delta_{i-2}}\rightarrow E_{i-1}^{1},{\mathop{\omega}\limits ^{\rightarrow}}_{i-2}\rightarrow{\mathop{\omega}\limits ^{\rightarrow}}_{i-1})=\frac{\mu_{T_1}(A_{i-1},E_{i-2}^{\delta_{i-2}}\rightarrow E_{i-1}^{1},{\mathop{\omega}\limits ^{\rightarrow}}_{i-2}\rightarrow{\mathop{\omega}\limits ^{\rightarrow}}_{i-1})}{\mu_{tot}(A_{i-1},E_{i-1}^{1})}
\end{eqnarray}
represent the PDFs of photons with energy $E_{i-2}^{\delta_{i-2}}$ moving in 
${\mathop{\omega}\limits ^{\rightarrow}}_{i-2}$, after undergoing Compton scattering or Rayleigh scattering at $A_{i-1}$, respectively, with respect to their scatter polar angles. If Compton scattering occurs at $A_{i-1}$, the energy changes and is denoted as $E_{i-1}^{0}$. If Rayleigh scattering occurs, the energy remains unchanged and is denoted as $E_{i-1}^{1}$.
	
In theory, after a photon reaches each $A_i$, 
it has a certain probability of reaching the detector. This paper calculates the probability of a photon reaching a fixed detector pixel at each $A_i$ to improve photon utilization. Thus
\begin{eqnarray}\label{f}
\fl	
		P(A_{i}
		\rightarrow D_{j})&=\sum_{\delta_{i}=0}^{1}p_{T_{\delta_i}}(A_{i})\int_{\Omega_{A_{i},D_{j} }}\mathrm{d}{\mathop{\omega}\limits ^{\rightarrow}}_{i,j}\\&\Big[p_{\theta_{\delta_i}}(A_i,E_{i-1}^{\delta_{i-1}}\rightarrow E_{i}^{\delta_{i}},{\mathop{\omega}\limits ^{\rightarrow}}_{i-1}\rightarrow {\mathop{\omega}\limits ^{\rightarrow}}_{i,j}){\rm exp}\big[{-\int_{0}^{b_{i,j}}\mu_{tot}(A_i+t{\mathop{\omega}\limits ^{\rightarrow}}_{i,j},E_i^{\delta_i})\mathrm{d}t}\big]\Big]\\&
		\approx
		\sum_{\delta_{i}=0}^{1}p_{T_{\delta_i}}(A_{i})p_{\theta_{\delta_i}}(A_i,E_{i-1}^{\delta_{i-1}}\rightarrow E_{i}^{\delta_{i}},{\mathop{\omega}\limits ^{\rightarrow}}_{i-1}\rightarrow {\mathop{\omega}\limits ^{\rightarrow}}_{i,j})\\&\times{\rm exp}\big[{-\int_{0}^{b_{i,j}}\mu_{tot}(A_i+t{\mathop{\omega}\limits ^{\rightarrow}}_{i,j},E_i^{\delta_i})\mathrm{d}t}\big]\Omega_{A_{i},D_{j} },
\end{eqnarray}
where $b_{ij}$ is the distance that a photon travels in M from $A_{i}$ along the forced scatter direction ${\mathop{\omega}\limits ^{\rightarrow}}_{i,j}$ to the detector pixel $D_j$. The $\Omega_{A_{i},D_{j}}$ is the solid angle corresponding to 
$D_j$ from the point $A_{i}$. $\Omega_{A_{i},D_{j}} \approx \frac{cos\alpha_{i,j} {h_{D_{j}}}^2}{|\overrightarrow{A_iD_{j}}|^2}, h_{D_{j}}^2$ is the area of the detector pixel $D_{j}$ and $\alpha_{i,j}$ represents the angle between ${\mathop{\omega}\limits ^{\rightarrow}}_{i,j}$ and the normal direction of 
	$D_{j}$.
	
From equations (1) - (4), and (7), it can be concluded that
\begin{eqnarray}\label{plij}
	\fl 
		P(l_{i,j})&=\sum_{\delta_{i}=0}^{1}p_{T_{\delta_i}}(A_{i})\int_{\Omega_{A_{i},D_{j} }}\mathrm{d}{\mathop{\omega}\limits ^{\rightarrow}}_{i,j}\\&\Big[p_{\theta_{\delta_i}}(A_i,E_{i-1}^{\delta_{i-1}}\rightarrow E_{i}^{\delta_{i}},{\mathop{\omega}\limits ^{\rightarrow}}_{i-1}\rightarrow {\mathop{\omega}\limits ^{\rightarrow}}_{i,j}){\rm exp}\big[{-\int_{0}^{b_{i,j}}\mu_{tot}(A_i+t{\mathop{\omega}\limits ^{\rightarrow}}_{i,j},E_i^{\delta_i})\mathrm{d}t}\big]\Big]\\& \prod_{k=2}^{i}\Big\{\sum_{\delta_{k-1}=0}^{1}p_{T_{\delta_{k-1}}}(A_{k-1})\int_{4\pi}\mathrm{d}{\mathop{\omega}\limits ^{\rightarrow}}_{k-1}\int_{0}^{c_{k-1}}\mathrm{d}t_{k}\\&\Big[p_{\theta_{\delta_{k-1}}}(A_{k-1},E_{k-2}^{\delta_{k-2}}\rightarrow E_{k-1}^{\delta_{k-1}},{\mathop{\omega}\limits ^{\rightarrow}}_{k-2}\rightarrow{\mathop{\omega}\limits ^{\rightarrow}}_{k-1})\\&\times \mu_{tot}(A_{k},E_{k-1}^{\delta_{k-1}}){\rm exp}\big[{-\int_{0}^{t_{k}}\mu_{tot}(A_{k-1}+t{\mathop{\omega}\limits ^{\rightarrow}}_{k-1},E_{k-1}^{\delta_{k-1}})\mathrm{d}t}\big]\Big]\Big\}\\&\times\int_{4\pi}\mathrm{d}{\mathop{\omega}\limits ^{\rightarrow}}_0\int_{\Omega_{E_0}}\mathrm{d}E_0\int_{0}^{c_0}\mathrm{d}t_1\Big[I_0({\mathop{\omega}\limits ^{\rightarrow}}_0)\phi(E_0)\\&\times \mu_{tot}(A_1,E_0){\rm exp}\big[{-\int_{0}^{t_1}\mu_{tot}(A_0+t{\mathop{\omega}\limits ^{\rightarrow}}_0,E_0)\mathrm{d}t}\big]\Big].
\end{eqnarray}

\subsection{Scatter Simulation Algorithm}

\begin{figure}[!t]
	\centerline{\includegraphics[width=\columnwidth]{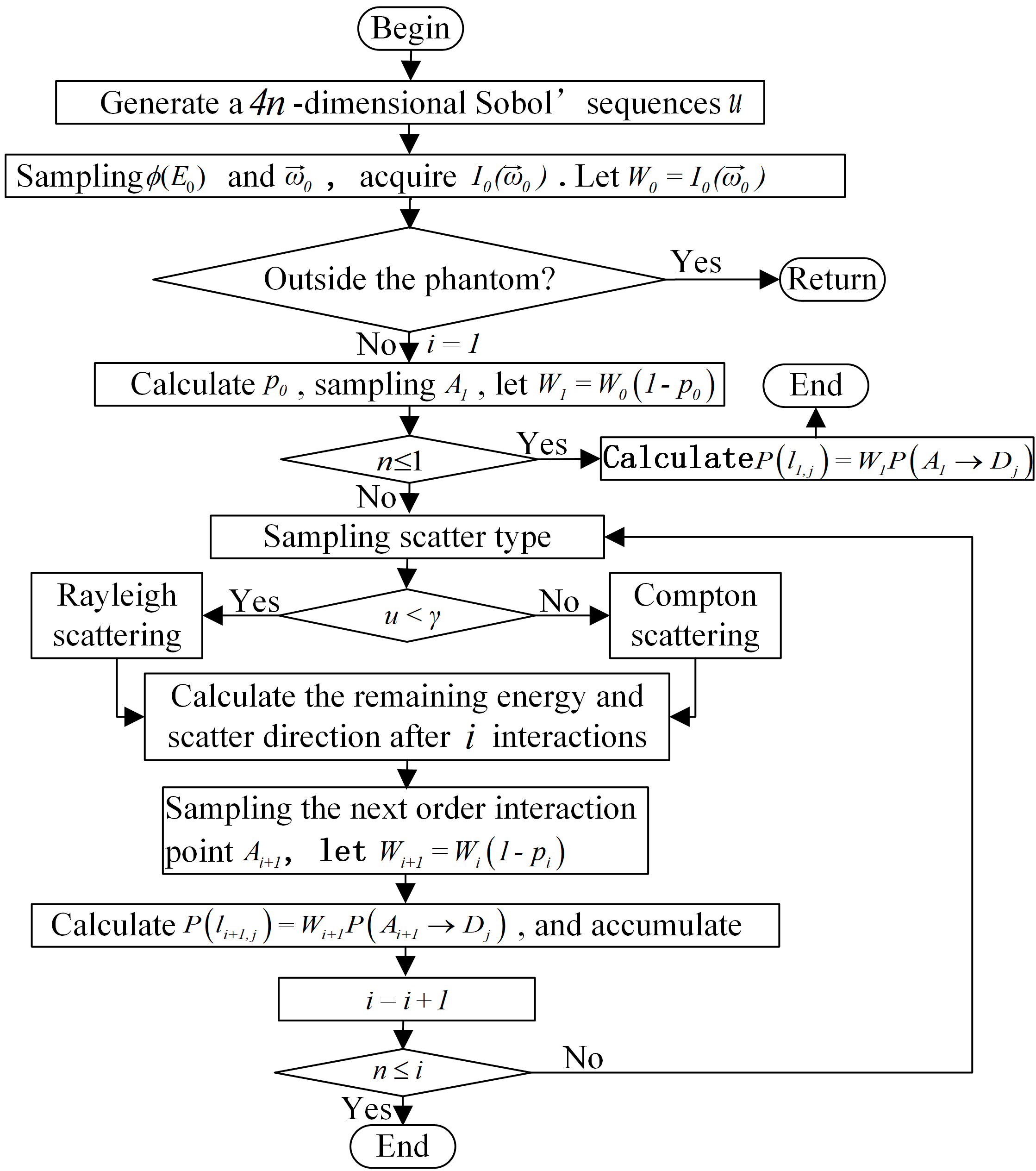}}
	\caption{An illustration of key steps in scatter simulation algorithm.}
	\label{fig:gQMCFFD202407-1.png}
\end{figure}

The intensity of the scatter signal from the measured object M received by the detector D after photons undergo $n$ interactions is 
\begin{equation}\label{total}
	\sum_{i=1}^{n}P(l_{i,j}), j=1,\dots,m^{2}.
\end{equation}
In this section, the algorithm for simulating equation (23) will be introduced, which employs QMC combined with forced fixed detection and down sampling. The specific steps are shown in figure 3.

Generate a $4n$-dimensional Sobol' points $\bi u=\{\bi u^{(k)}\},$ $k\in\{1,\cdots,N\}$ first, where $N$ is the number of simulations, $\bi u^{(k)}=\{u^{(k)}_{1},u^{(k)}_{2},\cdots,u^{(k)}_{4n}\}$. Sobol’ sequence is a type of low-discrepancy sequence\cite{On1967,2011Construction}. Using $u^{(k)}_{1}$ through Walker’s aliasing method\cite{Walker1977}, the initial energy $E_0$ is sampled from the energy spectrum $\phi(E)$. Use $u^{(k)}_{2}$ and $u^{(k)}_{3}$ to uniformly sample ${\mathop{\omega}\limits ^{\rightarrow}}_{0}$ in the initial illumination region, obtaining the initial flux $I_0(u^{(k)}_{2}, u^{(k)}_{3})$, let $W_0=I_0(u^{(k)}_{2}, u^{(k)}_{3})$. If the photon intersects with M, denote the initial intersection point as $A_0$. Then the photon moves within M along ${\mathop{\omega}\limits ^{\rightarrow}}_{0}$ with weight $W_0$. Calculate the probability $p_0({\mathop{\omega}\limits ^{\rightarrow}}_{0},E_0)$ that the photon escapes from M along ${\mathop{\omega}\limits ^{\rightarrow}}_{0}$ starting from $A_0$,
\begin{eqnarray}\label{p0}
	p_0({\mathop{\omega}\limits ^{\rightarrow}}_{0},E_0)={\rm exp}\Big[{-\int_{0}^{c_{0}}\mu_{tot}(A_{0}+t{\mathop{\omega}\limits ^{\rightarrow}}_{0},E_{0})\mathrm{d}t}\Big].
\end{eqnarray}
Sample the first-order interaction point $A_1$ according to the following formula
\begin{eqnarray}\label{u4}
	u^{(k)}_{4}=\frac{1-{\rm exp}\Big[{-\int_{0}^{t_1}\mu_{tot}(A_{0}+t{\mathop{\omega}\limits ^{\rightarrow}}_{0},E_{0})\mathrm{d}t}\Big]}{1-p_{0}({\mathop{\omega}\limits ^{\rightarrow}}_{0},E_0)}.
\end{eqnarray}
Let $W_1 = W_0(1-p_0({\mathop{\omega}\limits ^{\rightarrow}}_{0},E_0))$, $p(l_{1,j})=W_1P(A_{1}\rightarrow D_{j})$.

If $ n \leq 1$, then the calculation equation (23) is complete. Otherwise, use $u^{(k)}_{4i+1}$ to sample the scatter type in $A_i$ according to the weights $\gamma$ of interaction types combined with the importance sampling method. Use $u^{(k)}_{4i+2}$ and $u^{(k)}_{4i+3}$ to sample the $i$-th unit scatter direction ${\mathop{\omega}\limits ^{\rightarrow}}_{i}$ through the RITA\cite{2021Lin} algorithm. Similar to the $i = 0$ case, calculate the probability $p_{i}({\mathop{\omega}\limits ^{\rightarrow}}_{i},E_{i}^{\delta_{i}})$ of the photon escaping from M along ${\mathop{\omega}\limits ^{\rightarrow}}_{i}$ from $A_i$,
\begin{eqnarray}\label{pi}
	p_{i}({\mathop{\omega}\limits ^{\rightarrow}}_{i},E_{i}^{\delta_{i}})={\rm exp}\Big[{-\int_{0}^{c_{i}}\mu_{tot}(A_{i}+t{\mathop{\omega}\limits ^{\rightarrow}}_{i},E_{i}^{\delta_{i}})\mathrm{d}t}\Big].
\end{eqnarray}
Sample the next interaction point $A_{i+1}$ according to the following formula
\begin{eqnarray}\label{u4i+4}
	u^{(k)}_{4i+4}=\frac{1-{\rm exp}\Big[{-\int_{0}^{t_{i+1}}\mu_{tot}(A_{i}+t{\mathop{\omega}\limits ^{\rightarrow}}_{i},E_{i}^{\delta_{i}})\mathrm{d}t}\Big]}{1-p_{i}({\mathop{\omega}\limits ^{\rightarrow}}_{i},E_{i}^{\delta_{i}})}.
\end{eqnarray}
Let $W_{i+1} = W_i(1-p_{i}({\mathop{\omega}\limits ^{\rightarrow}}_{i},E_{i}^{\delta_{i}}))$, $$p(l_{i+1,j})=W_{i+1}P(A_{i+1}\rightarrow D_{j}).$$

Let $i = i + 1 $, if $n \leq i$, then the calculation equation (23) is complete. Otherwise, go back to sampling the scatter type and repeat the above steps until $n \leq i$.
\subsection{Spectrum Estimation Method with Scatter Correction}
The measured projection data will be affected by factors such as the energy spectrum, detector response, and scatter signal. The following model is used in the literature\cite{1997Ruth} to characterize the attenuation of X-rays in CT,
\begin{eqnarray}\label{5-3}
	I(l_j)=I_0(l_j)\Big[\int_{\Omega_E} \phi_d(E){\rm exp}\Big[-\int_{l_j}\mu_{tot}(\bi x, E)\mathrm{d} \bi x\Big]\mathrm{d} E+Sc_j\Big],
\end{eqnarray}
where  $I_0(l_j)$ is the initial flux intensity along the path $l_j$, $I(l_j)$ is the attenuation data along the path $l_j$, $\phi_d(E):=\phi(E)R(E)$ is the detected energy spectrum, $R(E)$ is the detector response function, $Sc_j$ is the intensity of scatter signal received by the detector pixel corresponding to path $l_j$. The equation (14) considers the influence of the energy spectrum, detector, and scatter signal on X-ray attenuation. To reconstruct the value or an approximation of $\mu_{tot}(\bi x, E)$, it is necessary to estimate $\phi_d(E)$ and $Sc_j$ in advance. This section proposes a spectrum estimation method with scatter correction based on the EM algorithm, referred to as EM-SC.

In addition to the scatter contamination caused by the measured object, the projection data obtained may also be affected by the environment, such as the detector housing. This paper first uses the SKS method to quickly simulate the scatter signal caused by the detector housing and performs an preliminary environmental scatter correction on $I_0(l_j)$ and $I(l_j)$, denoting the corrected  $I_0(l_j)$ and $I(l_j)$ as $I^*_0(l_j)$ and $I^*(l_j)$, respectively. Then use the scatter simulation algorithm proposed in this paper to simulate the scatter signal $Sc_j$ generated by the measured object. Therefore, the equation (28) can be converted into
\begin{eqnarray}\label{5-8}
		\hat{t}_j&:=\frac{I^*(l_j)}{I^*_0(l_j)}-Sc_j\\&=\int_{\Omega_E} \phi_d(E){\rm exp}\Big[-\int_{l_j}\mu_{tot}(\bi x, E)\mathrm{d} \bi x\Big]\mathrm{d} E.
\end{eqnarray}

Discretize equation (30) as follows
\begin{eqnarray}\label{5-9}
	\hat{t}_j=\sum_{i=1}^{I}K_{ij}w_i,\quad j\in J,
\end{eqnarray}
\begin{eqnarray}
	K_{ij}=\int_{\Omega_E}\varphi_i(E){\rm exp}\Big[-\int_{l_j}\mu_{tot}(\bi x, E)\mathrm{d} \bi x\Big]\mathrm{d} E, 
\end{eqnarray}
where $i\in \{1,\dots,I\}$, $I$ is the number of sampled spectra, $\varphi_i(E)$ is the selected energy spectrum basis, $w_i$ is the coefficient, 
\begin{eqnarray}\label{5-11}
	\phi_d(E)=\sum_{i=1}^{I}w_i\varphi_i(E).
\end{eqnarray}

By applying the EM algorithm iteration proposed by Sidky et al.\cite{2005EM} to equation (30) to solve for the desired spectrum, the following iterative solution form can be obtained:
\begin{eqnarray}
	w^{(k+1)}_i=\frac{w^{(k)}_i}{\sum_{j}K_{ij}}\sum_{j=1}^{J}K_{ij}\frac{\hat{t}_j}{\sum_{i}K_{i'j}w^{(k)}_{i'}},
\end{eqnarray}
where $k$ is the number of iterations. After the initial energy spectrum $\phi^{(0)}_d(E)$ is given, $w^{(0)}_i$ can be determined by equation (33).

\subsection{The Reconstruction Procedure}
In this section, we will provide detailed steps for the reconstruction, and the overall framework can be seen in figure 1.
\begin{enumerate}[(1)]

\item
Preparation phase, estimate the X-ray energy spectrum: at this stage, this paper uses a uniform aluminum (Al) phantom as the sample and collects attenuation data. First, the scatter signals caused by the detector housing in the attenuation datathe are corrected using the SKS method. Then the quantity $Sc_j$ is simulated using the scatter simulation algorithm proposed in this paper. Finally, the detection spectrum $\phi_d(E)$ of the X-rays is estimated by employed the EM-SC algorithm. This part corresponds to steps 1-3 in figure 1.

\item
Reconstruct the initial image of the measured object\label{Reconstruct}: collect the projection data of the measured object, such as the head, and the original volumetric data image is reconstructed using the FDK algorithm. This part corresponds to steps 4-5 in figure 1.

\item
Create material density volume and down sampling: segment the reconstructed CT images into air, water, and bone based on CT values, and down sampling the voxels. This part corresponds to step 6 in figure 1.

\item
Simulate the scatter signal of the measured object: based on the scatter model proposed in this paper, the scatter signal of the measured object is simulated using QMC combined with forced fixed detection and down sampling.

\item
Perform scatter signal and beam hardening correction on the projection data: first, perform preliminary scatter correction on the projection data using the SKS method to obtain projection data that has been corrected for scatter signals caused by the detector housing. Then, subtract the product of the scatter signal distribution caused by the measured object and a preset coefficient from the projection data, as detailed in the method proposed in reference\cite{2006zhu}. Next, perform beam hardening correction on the projection data using the spectrum curve of the detection spectrum $\phi_d(E)$. Finally, reconstruct and update the CT image using the FDK algorithm. This part corresponds to steps 7-8 in figure 1.

\item
Determine whether the CT image has converged: if the updated image shows little to no difference from the image reconstructed in the previous iteration, we consider that the image has converged. If converged, the image is denoised using the BM4D algorithm\cite{BM4D2013}, completing the CT image reconstruction; otherwise, return to 2) until the image converges. This part corresponds to steps 9-10 in figure 1.

\end{enumerate}

\section{Experiments and Results}
To evaluate the performance of the energy spectrum estimation algorithm EM-SC and the reconstruction method FDK-QMC-BM4D, 
an Al phantom (simulated data), an Al block (real data), the Shepp-Logan (SL) phantom \cite{shepp1974} (simulated data), and a head (real data) are used, respectively. In addition to visual consistency, we calculate the root mean square error (RMSE) to evaluate the performance of the EM-SC 
and the relative difference (RD), spatial non-uniformity (SNU), the universal quality index (UQI) \cite{2002AWang} and scatter to primary ratio (SPR) to quantify the quality of the reconstruction CT images. The quantities RD, SNU, UQI, and SPR are calculated as follows:

\begin{eqnarray}
	{\rm RD}=\frac{||r-t||_2}{||r||_2},
\end{eqnarray}
where $r$ is the reference value, $t$ is to be evaluated, $||\gamma||_2=\sqrt{\sum_{i=1}^{N}{\gamma_i}^2}$.

The quantity ${\rm SNU}$ is introduced to evaluate the CT images uniformity of each material
\begin{eqnarray}
	{\rm SNU}=\frac{{\mu}_{max}-{\mu}_{min}}{{\mu}_{wat}},
\end{eqnarray}
where ${\mu}_{min}$ and ${\mu}_{max}$ are the minimum mean and maximum values of the linear attenuation coefficient of the selected region of interest (ROI), respectively, ${\mu}_{wat}$ is the linear attenuation coefficient of water.   

The quantity UQI is proposed by Wang \emph{et al.} \cite{2002AWang}, which considers three factors of image reconstruction distortion: loss of correlation, luminance distortion and contrast distortion, and has been widely preformed to evaluate the quality of CT images. The definition of UQI is
\begin{eqnarray}
	{\rm UQI}=\frac{4\mathrm{Cov}(\bi x,\bi y)}{\mathrm{Var}(\bi x)+\mathrm{Var}(\bi y)}\times\frac{\bar{\bi x}\bar{\bi y}}{{\bar{\bi x}}^2+{\bar{\bi y}}^2},
\end{eqnarray}
where $\bi x=\{x_i|i=1,2,\dots,N\}$ is to be evaluated, $\bi y=\{y_i|i=1,2,\dots,N\}$ is the reference value. $\bar{\bi x}$ and $\mathrm{Var}(\bi x)$ are the mean and variance of $\bi x$, respectively. $\mathrm{Cov}(\bi x,\bi y)$ is the covariance of $\bar{\bi x}$ and $\bar{\bi y}$. The closer the UQI is to $1$, the better the effect of the method to be evaluated.

The quantity SPR is the scatter to primary ratio, and
\begin{eqnarray}
	{\rm SPR}=\frac{S_c}{S_p},
\end{eqnarray}
where $S_c$ is scatter signal intensity, $S_p$ is primary signal intensity. The larger ${\rm SPR}$, the greater the influence of the scatter signal on the reconstructed CT images. Besides, all experiments are performed in GPU (RTX3090). 

Here, we first provide an explanation of several reconstruction methods we will compare. The FDK-ECC method performs beam hardening on CT images based on the FDK, where ECC is a method for correcting beam hardening artifacts in CT\cite{ECC2006}. The FDK-ECC-BM4D method denoises the CT images using the BM4D method based on the FDK-ECC. The FDK-MC-GPU, FDK-SKS, and FDK-fASKS methods are based on the FDK, and they respectively use the MC-GPU, SKS, and fASKS to simulate scatter signals in CT. The FDK-MC-GPU-BM4D, FDK-ECC-BM4D, FDK-SKS-BM4D, and FDK-fASKS-BM4D methods denoise the CT images using the BM4D method based on their respective methods.

\subsection{Spectrum Estimation}

\begin{figure}[!t]
	\centerline{\includegraphics[width=\columnwidth]{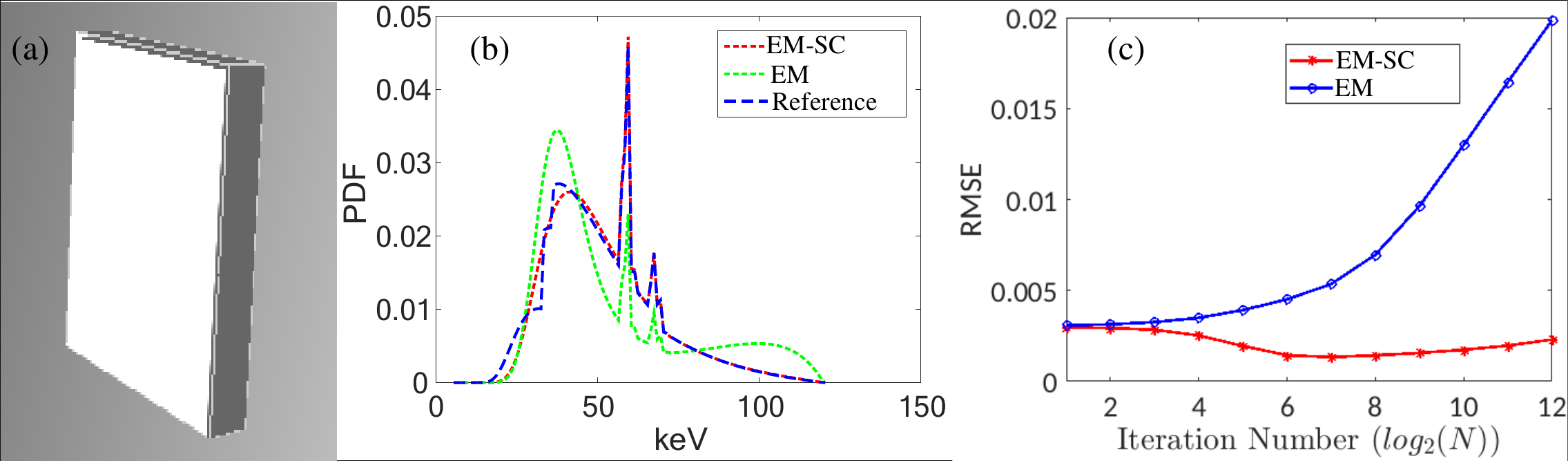}}
	\caption{Experimental results of the detected energy spectrum estimation for simulated data. (a) is the geometric illustration of the Al phantom; (b) shows the reference spectrum (blue line), the spectrum estimated by EM (green line) after 128 iterations, and the spectrum estimated by EM-SC (red line) after 128 iterations; (c) shows the RMSE of the spectra estimated by EM and EM-SC compared to the reference spectrum, with $N$ representing the number of iterations.
	}
	\label{fig:ESnoScattervswithScatter128RMSEnew}
\end{figure}

\subsubsection{Al phantom (simulated data)}
{We first validate the accuracy and robustness of the EM-SC algorithm using an Al phantom, and then use it to estimate the 
	$\phi_d(E)$ for the head experiment. 
	The Al phantom is obtained by segmenting and binarizing the reconstruction result of FDK and its volume is $80$ ${\rm mm}\times 80$ ${\rm mm}\times80$ ${\rm mm}$. The Al phantom is shown in figure 4(a). Rotating the Al phantom, the attenuation data of photons passing through it along different paths is simulated by MC-GPU \cite{Badal2009Accelerating}, and the 
	$Sc_j$ is simulated by the proposed scatter simulation algorithm. 

	The detected energy spectrums $\phi_d(E)$ of the Al phantom estimated by the EM-SC ($128$ iterations) and the EM ($128$ iterations) are presented in figure 4(b), and their RMSEs are $0.0013$ and $0.0054$, respectively. The figure and RMSE show that the energy spectrum estimated by the EM-SC is in good agreement with the reference energy spectrum. Figure 4(c) shows the RMSE of the EM and EM-SC algorithms, as well as their trend with respect to the number of iterations, respectively. It can be observed from figure 4(c) that the RMSE of the EM increases with the number of iterations, which may be caused by the introduction of noise during the iteration process. This indicates that the EM algorithm has weaker robustness for data contaminated with scatter signals. The RMSE of EM-SC initially decreases with the increase in the number of iterations, reaching a minimum at $128$ iterations, and remains stable with further iterations. This demonstrates the robustness of EM-SC.
	
	\begin{figure}[!t]
		\centerline{\includegraphics[width=\columnwidth]{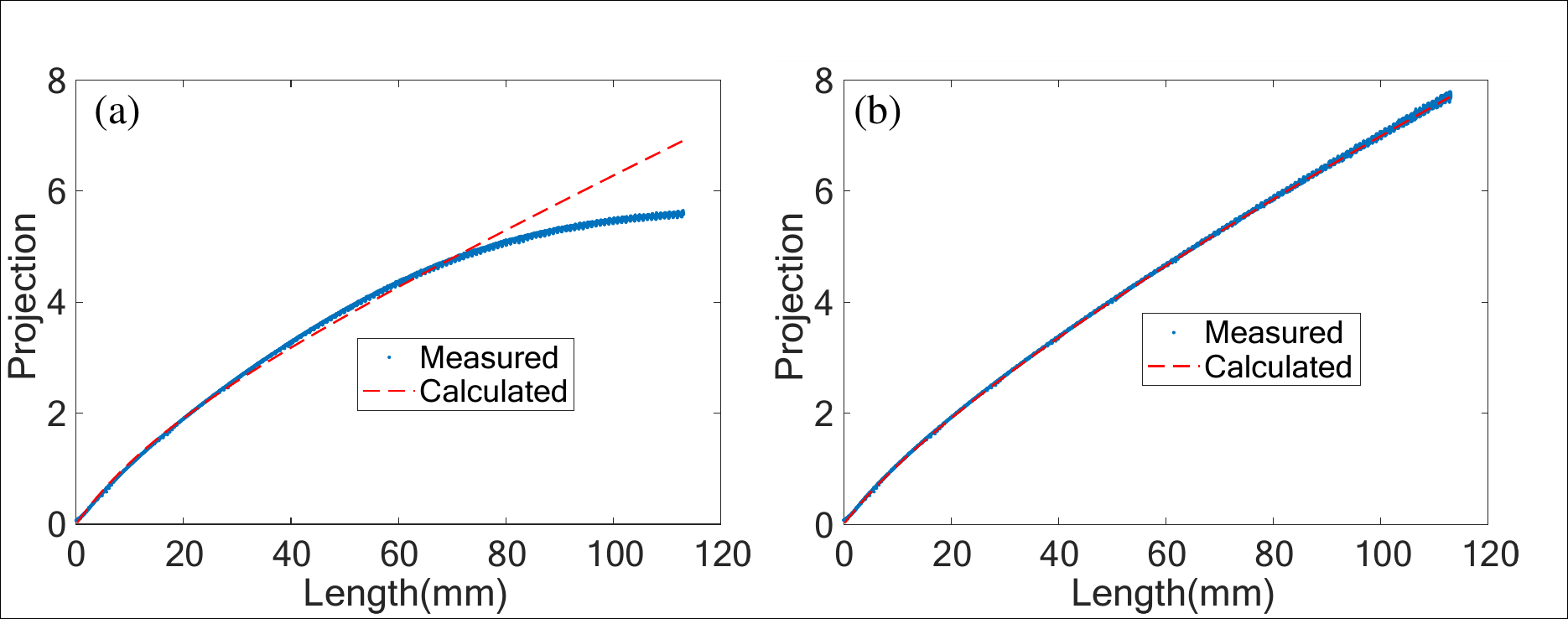}}
		\caption{The polychromatic projection curve of simulated data. (a) and (b) are the polychromatic projection curve of the detected energy spectrum estimated by EM and EM-SC, respectively.}
		\label{fig:ESnoScattervswithScatter128touyinzhinew}
	\end{figure}
	
	In order to further illustrate the accuracy of the EM-SC algorithm, we present a graph showing the relationship between the multi-energy projection values of the detected spectrum $\phi_d(E)$ and the chord length of the Al phantom, as shown in figure 5. Figure 5(a) and (b) show the relationships between the multi-energy projection values and the chord length of the Al phantom before and after scatter correction, respectively. The relationship graph composed of blue dots is obtained from the measured data, while the red dashed line is derived from the detected spectrum. Figure 5(a) shows that the scatter signal leads to a nonlinear decrease in the multi-color projection values, which becomes more pronounced as the chord length increases. This indicates the necessity of correcting the scatter signal and explains why EM-SC is more accurate than EM in handling data with scatter. In addition, it can be observed that the relationship curve between the multi-color projection values calculated using the spectrum estimated by EM-SC and the path length through the Al phantom fits closely with the measured relationship curve. This further demonstrates the accuracy of EM-SC.
} 

\subsection{Shepp-Logan Phantom}
\begin{figure}[!t]
	\centerline{\includegraphics[width=\columnwidth]{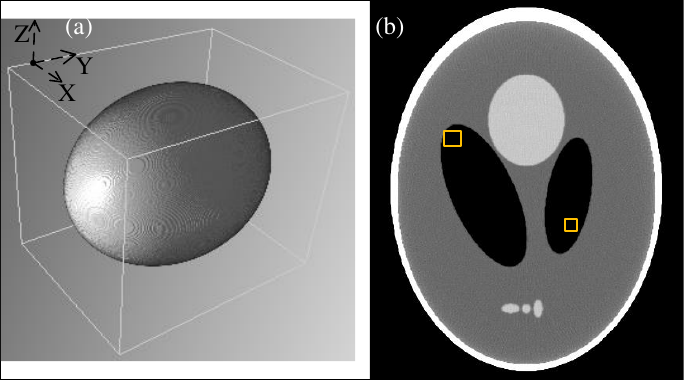}}
	\caption{(a) A geometric illustration of the SL phantom. (b) the transverse section of the SL, with ROIs marked by two yellow rectangles.}
	\label{fig:SLmodel5andIROnew}
\end{figure}

In this section, the SL phantom composed of aluminum, soft tissue and air are used to verify the effectiveness of the FDK-QMC-BM4D reconstruction method. During the scanning process, the phantom is in the field of view (IFOV), with the distances from the X-ray source S to the SL phantom and the detector being $500$ ${\rm mm}$ and $1000$ ${\rm mm}$, respectively. During the reconstruction process, the phantom is divided into $320\times 400\times 360$ voxels, and the voxels are down sampling to $80\times 100\times 90$, with each voxel having a size of $0.5$ ${\rm mm}\times0.5$ ${\rm mm}\times0.5$ ${\rm mm}$. Figure 7(a) shows the geometric illustration of the SL, while figure 7(b) is the transverse section of the SL, with ROIs marked by two yellow rectangles. 
\begin{table}[H]
	\caption{The RD, UQI, and SNP of different sections of the SL reconstructed by the FDK, FDK-SKS-BM4D, FDK-fASKS-BM4D, and FDK-QMC-BM4D methods, respectively. Here, $\#1$, $\#2$, and $\#3$ represent the transverse section, coronal section, sagittal section 
		of SL, respectively.} 
	\centering
	\begin{tabular}{ccccc}
		\toprule
		
		Metric&Methods &$\#1$& $\#2$& $\#3$\\
		\midrule
		\multirow{4}{*}{RD}&FDK&$10.76\%$&$9.36\%$&$9.36\%$\\
		&FDK-SKS-BM4D&$1.89\%$&$1.14\%$&$1.48\%$\\
		&FDK-fASKS-BM4D&$2.46\%$&$1.58\%$&$0.80\%$\\
		&FDK-QMC-BM4D&$1.55\%$&$0.80\%$&$1.14\%$\\
		\midrule
		\multirow{4}{*}{UQI}&FDK&$98.81\%$&$99.09\%$&$99.51\%$\\
		&FDK-SKS-BM4D&$99.97\%$&$99.98\%$&$99.98\%$\\
		&FDK-fASKS-BM4D&$99.95\%$&$99.98\%$&$99.94\%$\\
		&FDK-QMC-BM4D&$99.98\%$&$99.99\%$&$99.99\%$\\
		\midrule
		\multirow{4}{*}{SPR}&FDK&$2.23\%$&$6.17\%$&$6.28\%$\\
		&FDK-SKS-BM4D&$6.28\%$&$0.82\%$&$0.82\%$\\
		&FDK-fASKS-BM4D&$0.01\%$&$0.67\%$&$0.69\%$\\
		&FDK-QMC-BM4D&$0.08\%$&$0.36\%$&$0.24\%$\\		
		\bottomrule
	\end{tabular}
\end{table}
\begin{table}[H]
	\caption{The SNU of the SL reconstructed by the FDK and FDK-QMC-BM4D methods, respectively.}
	\centering
	\begin{tabular}{ccc}
		\toprule
		
		Metric&FDK&FDK-QMC-BM4D\\
		\midrule
		SNU&$35.83\%$&$5.976\%$\\
		\bottomrule
	\end{tabular}
\end{table}
\begin{figure}[!t]
	\label{fig:SLunscattervsuncorrectvscorrecttotalnew}
    \centerline{\includegraphics[width=\columnwidth]{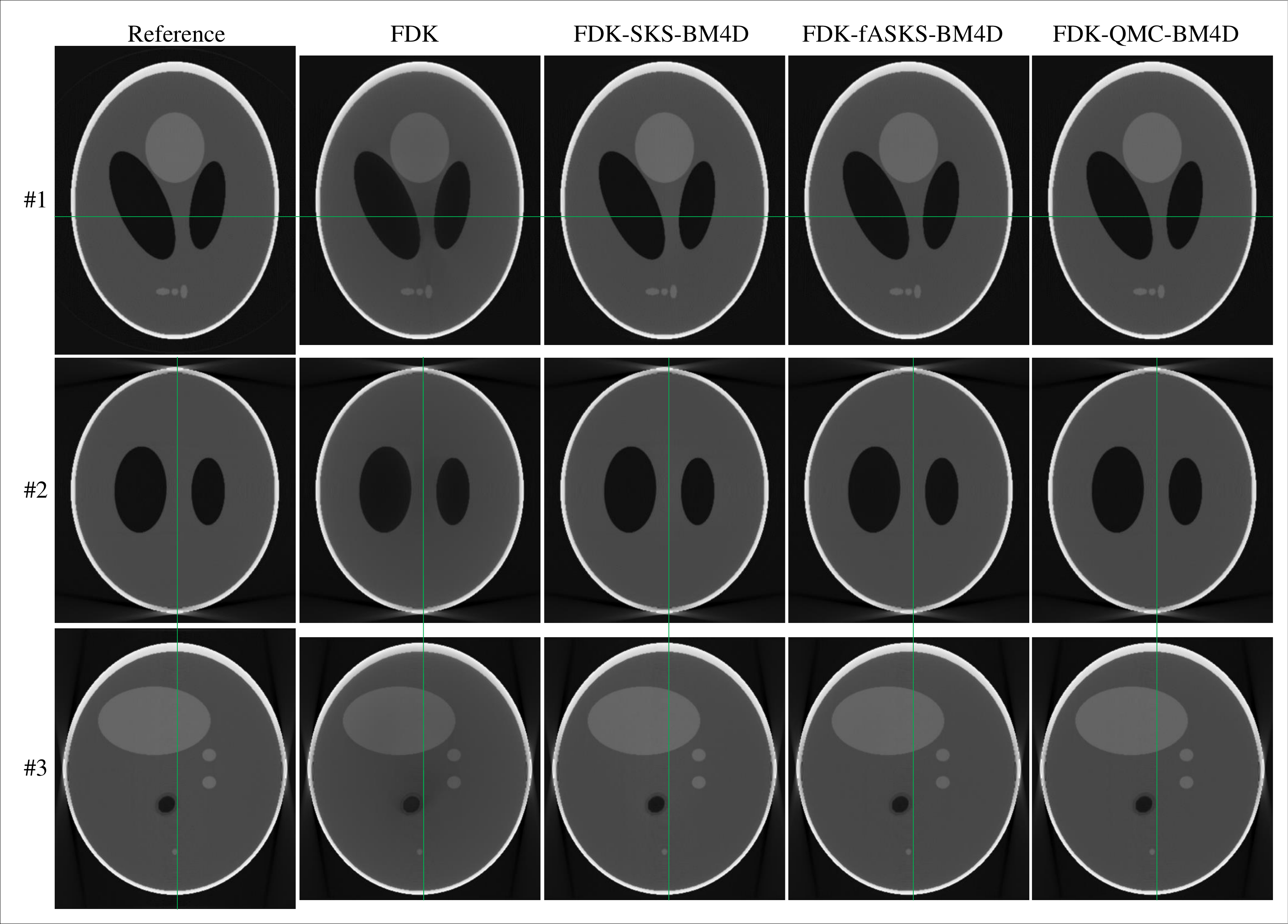}}
	\caption{CT images of the SL phantom reconstructed using the FDK, FDK-SKS-BM4D, FDK-fASKS-BM4D, and FDK-QMC-BM4D methods, respectively. $\#1$, $\#2$, and $\#3$ represent the transverse section, coronal section, sagittal section of SL, respectively.}
\end{figure}

Figure 8 shows the CT images of the SL phantom reconstructed using the FDK, FDK-SKS-BM4D, FDK-fASKS-BM4D, and FDK-QMC-BM4D methods, respectively. It can be seen that the CT images reconstructed using the FDK-QMC-BM4D method are visually highly consistent with the reference images, while the CT images reconstructed using the FDK-SKS-BM4D and FDK-fASKS-BM4D methods still contain some artifacts from figure 8. From figure 9, it can be observed that the profile of the FDK-QMC-BM4D reconstructed image fits highly with the profile of the reference image. This demonstrates the accuracy and robustness of the FDK-QMC-BM4D method. In addition, using the scatter simulation algorithm proposed in this paper, it takes approximately $2$ seconds to simulate the scatter signal image at each fixed angle, and about $1$ minute to obtain the converged CT image ($2$ iterations).

\begin{figure}[!t]
	\centerline{\includegraphics[width=\columnwidth]{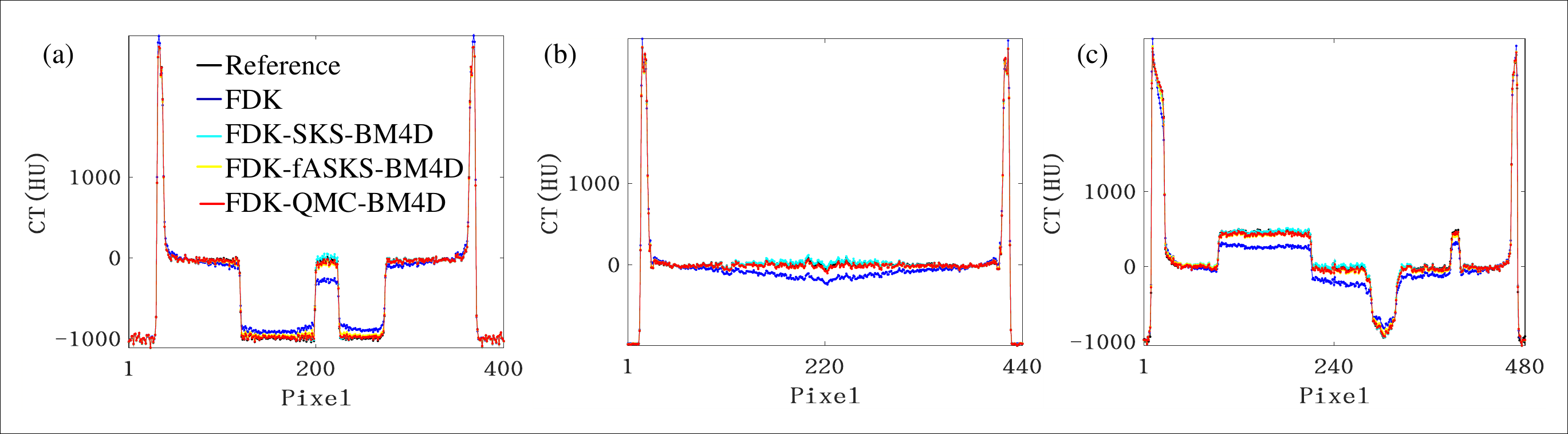}}
	\caption{(a), (b), and (c) are the profiles of green line of $\#1$, $\#2$, and $\#3$, respectively, in Fig. 8} 
\label{fig:SLreconrevsFDKvsSKSvsfASKSvsgQMCFFDline.pdf}
\end{figure}

Table 1 presents the numerical results of RD, UQI, and SPR for the transverse section, coronal section, sagittal section of SL reconstructed by the FDK, FDK-SKS, FDK-fASKS, and FDK-QMC methods. The RD values of the FDK-QMC method for different cross-sections are $1.55\%$, $0.80\%$, and $1.14\%$, which are lower than the results of the other methods; 
the UQI values are $99.98\%$, $99.99\%$, and $99.99\%$, which are very close to $1$; the SPR values are $0.08\%$, $0.36\%$, and $0.24\%$, respectively. Table 2 presents the SNU of FDK and FDK-QMC-BM4D, which are $35.83\%$ and $5.976\%$, respectively. These data further quantify and illustrate the accuracy and robustness of the FDK-QMC-BM4D method.

\subsection{Head}

In this section, we use a head to verify the effectiveness of the FDK-QMC-BM4D method. The schematic diagram of the projection data acquisition can be found in figure 1 under ``Scanner''. During the scanning process, the head is IFOV, with the distances from the X-ray source S to the head and the detector being $740$ ${\rm mm}$ and $1125$ ${\rm mm}$, respectively. During the reconstruction process, the head is divided into $864\times752\times1024$ voxels, and the voxels are down sampling to $108 \times94\times128$, with each voxel measuring $0.263$ ${\rm mm} \times 0.263$ ${\rm mm} \times 0.263$ ${\rm mm}$.

\begin{figure}[!t]
	\label{fig:Head80keVXZ342line}
	\centering
	\includegraphics[scale=0.17]{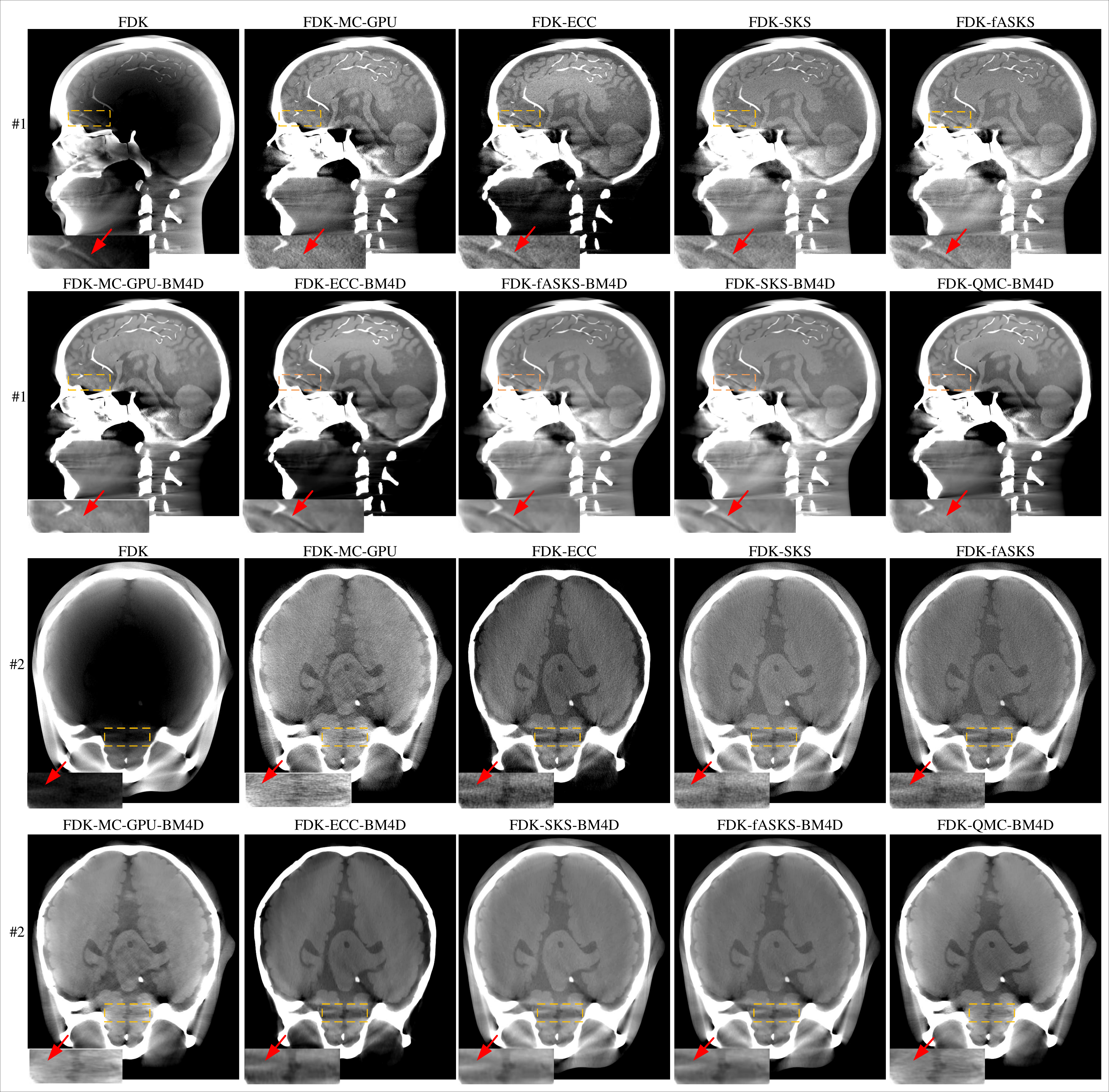}
	\caption{CT images of the head reconstruced using the FDK, FDK-MC-GPU(-BM4D), FDK-ECC(-BM4D), FDK-SKS(-BM4D), FDK-fASKS(-BM4D), and FDK-QMC-BM4D methods, respectively. The image pointed to by the red arrow is an enlargement of the ROI. $\#1$ and $\#2$ are different cross-sections of the head. The reconstruction results of the head with energy range of $0$ to $80$ keV. Display window $[-200,200]$ HU. 
	}
\end{figure}

Figure 10 shows the head CT images reconstructed by the FDK, FDK-MC-GPU(-BM4D), FDK-ECC(-BM4D), FDK-SKS(-BM4D), FDK-fASKS(-BM4D), and FDK-QMC-BM4D methods, respectively. 
From figure 10, it can be seen that the CT images reconstructed by the FDK-QMC-BM4D method are highly consistent with those reconstructed by the FDK-MC-GPU-BM4D method. The ROIs are uniform and smooth, with no dents. FDK-QMC-BM4D and FDK-MC-GPU(-BM4D) are consistent with the expected results, while the other methods exhibit dents and non-uniformities. The entire reconstruction process takes approximately $1.25$ minutes with the FDK-QMC-BM4D method, whereas the FDK-MC-GPU-BM4D method takes about $128$ minutes. CT images reconstructed using the FDK-ECC(-BM4D), FDK-SKS(-BM4D), and FDK-fASKS(-BM4D) methods exhibit a certain amount of residual artifacts. 
Furthermore, the results in figure 10 also demonstrate that denoising the images is necessary.

\section{Discussion}
The scatter model established in this paper accurately characterizes the interaction between photons and matter, and the proposed scatter simulation algorithm can effectively and rapidly simulate the scatter signal in CT. In our experiments, the scatter simulation algorithm takes approximately $2$ to $2.5$ seconds to simulate the scatter signals at fixed angles, whereas the MC-GPU algorithm takes about $127$ seconds. However, the complexity of this scatter simulation algorithm greatly increases as the scatter order $n$ increases. If random sequences are used instead of low-discrepancy sequences to simulate scatter signals in the algorithm, more photons will be require to achieve the same quality of scatter signal images, leading to an increase in runtime.

In the simulated data, the image reconstructed using the FDK-QMC-BM4D method highly fits the reference image, whereas the images reconstructed using the FDK-SKS-BM4D and FDK-fASKS-BM4D methods contain residual artifacts. Moreover, the numerical results of image quality metrics RD, UQI, SPR, and SNU also demonstrate the effectiveness of FDK-QMC-BM4D. In the head experiment, compared with the FDK method 
and the FDK-ECC(-BM4D) method with only beam hardening correction, the images reconstructed using the FDK-QMC-BM4D method have almost no residual artifacts and the image quality is significantly improved. Furthermore, the results reconstructed by these three methods further elucidate the impact of scatter signals and beam hardening on image quality, highlighting the importance of correcting scatter signals and beam hardening in CT. Compared to the FDK-SKS(-BM4D) and FDK-fASKS(-BM4D) methods, the CT images reconstructed using the FDK-QMC-BM4D method exhibit stronger contrast and no residual artifacts in the ROIs. Compared to the FDK-MC-GPU(-BM4D) method, the FDK-QMC-BM4D method improves the running speed by approximately $102$ times while ensuring accuracy. This makes it possible to apply the FDK-QMC-BM4D method in clinical CT reconstruction.

In addition, the scatter model and scatter simulation algorithm proposed in this paper are also applicable to the simulation of scatter signals in spectral CT and grating-based phase-contrast imaging. Does it have equally good effects on spectral CT and grating-based phase-contrast imaging? The real experiments conducted in this paper were performed on a head. Does the FDK-QMC-BM4D method achieve similarly good results for industrial CT samples containing metal, such as circuit boards, blades, etc.?  We will further study and refine these work in the future.
\section{Conclusion}
This paper presents a multi-module coupled reconstruction method, FDK-QMC-BM4D, which simultaneously eliminates scatter artifacts, beam hardening artifacts, and noise in CT images. It effectively reduces artifacts in CT images and improves image contrast and resolution. Compared to the widely recognized MC method (MC-GPU), which is known for its precise simulation of particle transport, this method can reduce the overall reconstruction time by tens to hundreds of times while achieving the same or even higher accuracy. However, the MC method, due to its enormous computational demands, has not yet been applied in clinical practice. The FDK-QMC-BM4D method is based on QMC simulation of scatter signals, where the QMC method is a deterministic variant of the MC method. Its fundamental idea is to use low-discrepancy sequences instead of the random numbers for simulation. Based on the comparison results with the MC method and the mechanism of the FDK-QMC-BM4D method, this study provides a new approach to addressing artifacts in clinical CT imaging.

\section*{Acknowledgment}
This work was supported in part by the National Natural Science Foundation of China under Grant 72071119, and Grant 12331011, in part by the National Key Research and Development Program of China under Grant 2020YFA0712200 and in part by the hunan provincial department of education scientific research project under Grant 24B0582. The authors also thank Beijing Advanced Innovation Center for Imaging Technology, Capital Normal University for assistance provided for the realization of the physical experiments.

\section*{References}

\bibliographystyle{jphysicsB}

\bibliography{harvard}

@article{2006zhu,
	author = {Zhu, L. and Bennett, N. R. and Fahrig, R.},
	title = {Scatter correction method for {X}-ray {C}{T} using primary modulation: theory and preliminary results},
	journal = {IEEE Trans. Med. Imaging},
	year    = {2006},
	volume  = {25},
	number  = {12},
	pages   = {1573–-1587},
}

@article{2015Ritschl,
	author = {Ritschl, L. and  Fahrig, R. and Knaup, M. and  Maier, J. and Kachelrieß, M.},
	title = {Robust primary modulation-based scatter estimation for cone-beam {C}{T}},
	journal = {Med.Phys.},
	year    = {2015},
	volume  = {42},
	number  = {1},
	pages   = {469-–478},
}

@article{2017Bier,
	author = {Bier, B. and Berger, M. and Maier, A. and  Kachelrieß, M. and Ritschl, L. and M\"uller, K. and Choi, J. H. and  Fahrig, R.},
	title = {Scatter correction using a primary modulator on aclinical angiography {C}-arm {C}{T} system},
	journal = {Med.Phys.},
	year    = {2017},
	volume  = {44},
	number  = {9},
	pages   = {e125–-e137},
}

@article{grid,
	author = { Kyriakou, Y. and Kalender, W.},
	title = {Efficiency of antiscatter grids for flat-detector {C}{T}},
	journal = {Phys. Med. Biol.},
	year    = {2007},
	volume  = {52},
	pages   = {6275--6293},
}

@article{2018CS,
	author = {Yang, F. and Zhang, D. and  Huang, K. and Shi, W. and Wang, X.},
	title = {Scattering estimation for cone-beam {C}{T} using local measurement based on compressed sensing},
	journal = {IEEE Trans. Nuclear Science},
	year    = {2018},
	volume  = {65},
	number  = {3},
	pages   = {941--949},
}

@article{SKS1988,
	author = {Seibert, J. A.  and Boone, J. M.},
	title = {X‐ray scatter removal by deconvolution},
	journal = {Med. Phys.},
	year    = {1988},
	volume  = {15},
	number  = {4},
	pages   = {567--575},
}

@article{SKS2010,
	author = {Sun, M.  and  Star-Lack, J. M.},
	title = {Improved scatter correction using adaptive scatter kernel superposition},
	journal = {Phys. Med. Biol.},
	year    = {2010},
	volume  = {55},
	pages   = {6695--6720},
}

@article{2004Colijn,
	author = {A. P. Colijn, and F. J. Beekman,},
	title = {Accelerated simulation of cone beam {X}-ray scatter projections},
	journal = {IEEE Trans. Med. Imaging},
	year    = {2004},
	volume  = {23},
	number  = {5},
	pages   = {584--590},
}

@article{2006Zbijewski,
	author = {W. Zbijewski, and F. J. Beekman,},
	title = {Efficient {M}onte {C}arlo based scatter artifact reduction in cone-beam micro-{C}{T}},
	journal = {IEEE Trans. Med. Imaging},
	year    = {2006},
	volume  = {25},
	number  = {7},
	pages   = {817--827},
}

@article{2006Kyriakou,
	author = {Y. Kyriakou, and T. Riedel, and W. A. Kalender,},
	title = {Combining deterministic and {M}onte {C}arlo calculations for fast estimation of scatter intensities in {C}{T}},
	journal = {Phys. Med. Biol.},
	year    = {2006},
	volume  = {51},
	number  = {18},
	pages   = {4567--4586},
}

@article{2009Poludniowski,
	author = {G. Poludniowski, and  P. M. Evans, and V. N. Hansen, and S. Webb,},
	title = {An efficient {M}onte {C}arlo-based algorithm for scatter correction in ke{V} cone-beam {C}{T}},
	journal = {Phys. Med. Biol.},
	year    = {2009},
	volume  = {54},
	number  = {12},
	pages   = {3847--3864},
}

@article{Badal2009Accelerating,
	author = {J. Bar\'o, and J. Sempau, and J. M. Fern\'andez-Varea, and F. Salvat, },
	title = {Accelerating {M}onte {C}arlo simulations of photon transport in a voxelized geometry using a massively parallel graphics processing unit},
	journal = {Med. Phys.},
	year    = {2009},
	volume  = {36},
	number  = {11},
	pages   = {4878--4880},
}

@article{JiaFast,
	author = {Jia, X. and Yan, H. and Gu, X. and Jiang, S. B.},
	title = {Fast {M}onte {C}arlo simulation for patient-specific {C}{T}/{C}{B}{C}{T} imaging dose calculation},
	journal = {Phys. Med. Biol.},
	year    = {2012a},
	volume  = {57},
	number  = {3},
	pages   = {577--590},
}

@article{2012A,
	author = {X. Jia, and H. Yan, and L. Cervi\~no, and M. Folkerts, and S. B. Jiang,},
	title = {A {G}{P}{U} tool for efficient, accurate, and realistic simulation of cone beam {C}{T} projections},
	journal = {Med. Phys.},
	year    = {2012b},
	volume  = {39},
	number  = {12},
	pages   = {7368--7378},
}

@article{gMMC2019,
	author = {Y. Xu, and Y. Chen, and  Z. Tian, and X. Jia, and L. H. Zhou,},
	title = {Metropolis {M}onte {C}arlo simulation scheme for fast scattered {X}-ray photon calculation in {C}{T}},
	journal = {Opt. Express},
	year    = {2019},
	volume  = {27},
	number  = {2},
	pages   = {1262--1275},
}

@article{gMMC2020,
	author = {Zhang, Y. and Chen, Y. and  Zhong, A. and Jia, X. and Wu, Y. and Qi, H. and Zhou, L. and Xu, Y.},
	title = {Scatter correction based on adaptive photon path-based Monte Carlo simulation method in Multi-GPU platform},
	journal = {Computer Methods and Programs in Biomedicine},
	year    = {2020},
	volume  = {194},
	pages   = {105487},
}

@article{2024dark,
	author = {T. Urban, and W. Noichl, and  K. J. Engel, and T. Koehler, and F. Pfeiffer, },
	title = {Correction for {X}-ray scatter and detector crosstalk in dark-field radiography},
	journal = {IEEE Trans. Med. Imaging},
	year    = {2024},
	volume  = {43},
	number  = {7},
	pages   = {2646--2656},
}

@article{2018Acuros,
	author = {Maslowski, A. and A. Wang, and M. Sun, and T. Wareing, and I. Davis, and J. Star-Lack,},
	title = {Acuros {C}{T}{S}: {A} fast, linear {B}oltzmann transport equation solver for computed tomography scatter - {P}art {I}: {C}ore algorithms and validation},
	journal = {Med. Phys.},
	year    = {2018a},
	volume  = {45},
	number  = {5},
	pages   = {1899--1913},
}

@article{2018Acurosb,
	author = {A. Maslowski, and A. Maslowski, and  P. Messmer, and M. Lehmann, and A. Strzelecki, and  E. Yu, and P. Paysan, and M. Brehm, and P. Munro, and J. Star-Lack, and D. Seghers,},
	title = {Acuros {C}{T}{S}: A fast, linear {B}oltzmann transport equation solver for computed tomography scatter - {P}art {I}{I}: {S}ystem modeling, scatter correction, and optimization},
	journal = {Med. Phys.},
	year    = {2018b},
	volume  = {45},
	number  = {5},
	pages   = {1914--1925},
}

@article{2019Scatter1,
	author = {Y. Jiang, and  C. Yang, and P. Yang, and X. Hu, and C. Luo, and Y. Xue, and X. Hu, and L. Zhang, and J. Wang, and T. Niu,},
	title = {Scatter correction of cone-beam {C}{T} using a deep residual convolution neural network ({D}{R}{C}{N}{N})},
	journal = {Phys. Med. Biol.},
	year    = {2019},
	volume  = {64},
	number  = {14},
	pages   = {145003},
}

@inproceedings{2020Roser,
	author    = {P. Roser, and X. Zhong, and A. Birkhold, and A. Preuhs, and C. Syben, and E. Hoppe, and N. Strobel, and M. Kowarschik, and R. Fahrig, and A. Maier,},
	title     = {Simultaneous estimation of x-ray back-scatter and forward-scatter using multi-task learning},
	booktitle = {International Conference on Medical Image Computing and Computer-Assisted Intervention},
	year      = {2020},
	pages     = {199–-208},
	publisher = {Springer},
}

@article{2021ScatterDL,
	author = {P. Roser, and Annette Birkhold, and Alexander Preuhs, and Christopher Syben, and Lina Felsner, and Elisabeth Hoppe, and Norbert Strobel, and Markus Korwarschik, and Rebecca Fahrig, and Andreas Maier,},
	title = {X-ray scatter estimation using deep splines},
	journal = {IEEE Trans. Med. Imaging},
	year    = {2021},
	volume  = {40},
	number  = {9},
	pages   = {2272--2283},
}

@article{2021ScatterDLHe,
	author = { J. He, and S. Chen, and H. Zhang, and X. Tao, and W. Lin, and S. Zhang, and D. Zeng, and J. Ma,},
	title = {Downsampled imaging geometric modeling for accurate {C}{T} reconstruction via deep learning},
	journal = {IEEE Trans. Med. Imaging},
	year    = {2021},
	volume  = {40},
	number  = {11},
	pages   = {2976--2985},
}

@book{niederreiter:1992,
	author    = {H. Niederreiter},
	title     = {Random {N}umber {G}eneration and {Q}uasi-{M}onte {C}arlo {M}ethods},
	publisher = {Society for Industrial and Applied Mathematics},
	year      = {1992},
	address   = {Philadelphia},
}

@book{dick:pillichshammer:2010,
	author    = {J. Dick, and F. Pillichshammer,},
	title     = {Digital {N}ets and {S}equences: {D}iscrepancy {T}heory and {Q}uasi-{M}onte {C}arlo {I}ntegration},
	publisher = {Cambridge University Press},
	year      = {2010},
}

@article{2021Lin,
	author = {G. Lin, and S. Deng, and X. Wang,},
	title = {Quasi-{M}onte {C}arlo method for calculating {X}-ray scatter in {C}{T}},
	journal = {Opt. Express},
	year    = {2021},
	volume  = {29},
	number  = {9},
	pages   = {13746--13763},
}

@article{2008Acceleration,
	author = {J. De Beenhouwer, and S. Staelens, and S. Vandenberghe, and I. Lemahieu,},
	title = {Acceleration of {G}{A}{T}{E} {S}{P}{E}{C}{T} simulations},
	journal = {Med. Phys.},
	year    = {2008},
	volume  = {35},
	number  = {4},
	pages   = {1476--1485},
}

@article{2005EM,
	author = {E. Y. Sidky, and  L. Yu, and X. Pan, and Y. Zou, and M. Vannier,},
	title = {A robust method of x-ray source spectrum estimation from transmission measurements: {D}emonstrated on computer simulated, scatter-free transmission data},
	journal = {Journal of Applied Physics},
	year    = {2005},
	volume  = {97},
	number  = {12},
	pages   = {1247},
}

@article{2015Lee,
	author = {J. S. Lee, and J. C. Chen,},
	title = {A single scatter model for {X}-ray {C}{T} energy spectrum estimation and polychromatic reconstruction},
	journal = {IEEE Trans. Med. Imaging},
	year    = {2015},
	volume  = {34},
	number  = {6},
	pages   = {1403--1413},
}

@article{1984FDK,
	author = {L. A. Feldkamp, and L. C. Davis, and J. W. Kress,},
	title = {Practical cone-beam algorithm},
	journal = {Journal of the Optical Society of America A},
	year    = {1984},
	volume  = {1},
	number  = {6},
	pages   = { 612--619},
}

@article{BM4D2013,
	author = {M. Maggioni, and V. Katkovnik, and K. Egiazaran, and A. Foi,},
	title = {Nonlocal transform-domain filter for volumetric data denoising and reconstruction},
	journal = {IEEE Trans. on Image Processing},
	year    = {2013},
	volume  = {22},
	number  = {1},
	pages   = {119--133},
}

@article{On1967,
	author = { I. M. Sobol',},
	title = {On the distribution of points in a cube and the approximate evaluation of integrals},
	journal = {Zh. Vychisl. Mat. Mat. Fiz.},
	year    = {2008},
	volume  = {7},
	number  = {4},
	pages   = { 86--112},
}

@article{2011Construction,
	author = { I. M. Sobol', and D. Asotsky, and A. Kreinin, and S. Kucherenko,},
	title = {Construction and comparison of high-dimensional {S}obol' generators},
	journal = {Wilmott},
	year    = {2011},
	volume  = {2011},
	number  = {56},
	pages   = {64--79},
}

@article{Walker1977,
	author = {A. J. Walker,},
	title = {An efficient method for generating discrete random variables with general distributions},
	journal = {ACM Trans. on Mathematical Software},
	year    = {1977},
	volume  = {3},
	number  = {3},
	pages   = {253-256},
}

@article{1997Ruth,
	author = {C. Ruth, and P. M. Joseph,},
	title = {Estimation of a photon energy spectrum for a computed tomography scanner},
	journal = {Med. Phys.},
	year    = {1997},
	volume  = {24},
	number  = {5},
	pages   = {695--702},
}

@article{shepp1974,
	author = {L. A. Shepp, and B. F. Logan,},
	title = {Reconstructing interior head tissue from {X}-ray transmissions},
	journal = {IEEE Transactions on Nuclear Science},
	year    = {1974},
	volume  = {21},
	number  = {1},
	pages   = {228--236},
}

@article{2002AWang,
	author = {Z. Wang, and A. C. Bovik,},
	title = {A universal image quality index},
	journal = {IEEE Signal Processing Letters},
	year    = {2002},
	volume  = {9},
	number  = {3},
	pages   = {81--84},
}

@article{ECC2006,
	author = { M. Kachelrieß, K. Sourbelle, and W. A. Kalender,},
	title = {Empirical cupping correction: A first-order raw data precorrection for cone-beam computed tomography},
	journal = {Med. Phys.},
	year    = {2006},
	volume  = {33},
	number  = {5},
	pages   = {1269--1274},
}

\end{document}